\documentclass[useAMS,usenatbib,usedcolumn,usegraphicx]{mn2e}

\def\etal{{\it et\thinspace al.}\ }
\def\o3{{[{\sc O\,iii}]}}

\def\lmlm{{$\lambda\lambda$}\ }
\def\mum{{$\mu$m}\ }

\def\eion{{(e~+~ion)}\ }
\def\te{{T$_e$}}
\def\ne{{N$_e$}}

\usepackage[dvips]{graphics}
\usepackage[dvipsnames,usenames]{color}

\newcommand{\be}{\begin{equation}}
\newcommand{\ee}{\end{equation}}

\title{{Improved collision strengths and line ratios
for forbidden [{\bf\sc O\,iii}] far-infrared and optical lines}}
\author
[Ethan Palay, Sultana N. Nahar, Anil K.\ Pradhan]
       {Ethan Palay$^1$, Sultana N. Nahar$^1$, Anil K. Pradhan$^1$,
Werner Eissner$^2$\\ 
       $^1$ Department of Astronomy,
 The Ohio State University, Columbus, OH 43210, USA,\\$^2$ Institut
 f\"ur Theoretische Physik, Teilinstitut 1, 70550 Stuttgart, Germany}
\date{Accepted  xxxxxx 
      Received xxxxxx;
      in original form xxxxxx}

\pagerange{\pageref{firstpage}--\pageref{lastpage}}
\pubyear{2011}

\def\LaTeX{L\kern-.36em\raise.3ex\hbox{a}\kern-.15em
    T\kern-.1667em\lower.7ex\hbox{E}\kern-.125emX}

\begin{document}

\maketitle

\label{firstpage}

\begin{abstract}
 Far-infrared and optical [{\sc O\,iii}] lines are useful temeprature-density
diagnostics of nebular as well as dust obscured astrophysical sources.
Fine structure transitions among the ground state levels $1s^22s^22p^3 \
^3P_{0,1,2}$ give rise to the 52 and 88 $\mu$m lines, whereas
transitions among the $^3P_{0,1,2}, ,^1D_2, ^1S_0$ levels yield the
well-known optical lines \lmlm 4363, 4959 and 5007 $\AA$.
These lines are excited primarily by electron impact excitation.
But despite their importance in nebular diagnostics collision strengths
for the associated fine structure transitions have not been computed
taking full account of relativistic effects. We present 
Breit-Pauli R-matrix calculations for the collision strengths with 
highly resolved
resonance structures. We find significant differences
of up to 20\% in the Maxwellian averaged rate coefficients from previous
works. We also tabulate these to lower temperatures down to 100 K to
enable determination of physical conditions in cold dusty environments
such photo-dissociation regions and
ultra-luminous infrared galaxies observed with the {\it
Herschel} space observatory.
We also examine the effect of improved collision strengths on 
temperature and density sensitive line ratios.
\end{abstract}

\begin{keywords}
Gaseous Nebulae -- Optical Spectra: {\sc H\,ii} Regions -- Line Ratios: 
Atomic Processes -- Atomic Data
\end{keywords}

\section{Introduction}

 \o3 optical lines have long been standard nebular temperature diagnostics
with wavelengths almost in the middle of the optical spectrum at
\lmlm 4363, 4959, 5007 (viz. Aller 1956, Dopita and Sutherland 2003,
Pradhan and Nahar 2011).
In recent years, owing to the advent of far-infrared (FIR) space
observatories and instruments such as the {\it Intfrared Space
Observatory} -
Long Wavelength Spectrograph (ISO-LWS), the {\it Spitzer} Infrared
Spectrograph, and the {\it Herschel} Photodetector Array Camera and
Spectrometer (PACS) the \o3 FIR lines have proven to have great
potential in providing diagnostics of physical conditions in a variety
of astrophysical objects that are generally obscured by dust extinction
at optical or shorter wavelengths. These range from Galactic {\sc H\,ii}
regions (Martin-Hernandez \etal 2002, Morisset \etal 2002) to
star-forming galaxies at intermediate redhift (Liu \etal 2008) 
and ultra-luminous infrared galaxies (ULIRGs). 
For example, the
\o3 FIR lines at \lmlm 88 and 52 \mum are observed from dusty ULIRGs,
which are copious IR emitters and become
more prominent with increasing redshift (Nagao \etal 2011). They
 may be valuable indicators of the
metallicity evolution from otherwise inaccessible star-forming regions
buried deep within the galaxies (Houck \etal 2004, 2005). 
 
 The forbidden FIR lines arise from very low-energy excitations 
within the fine structure
levels of the ground state of atomic ions, such as the \o3 $^3P_o
\rightarrow ^3P_1$ transition at 88.36 \mum and the $^3P_1
\rightarrow ^3P_2$ transition at 51.81 \mum. As such they can be excited
by electron impact at low temperatures, even \te $\sim$ 1000 K or less.
That also accounts for their utility since the FIR lines can be formed
in (and therefore probe) not
only {\sc H~ii} regions but also photo-dissociation regions (PDRs) where the
temperaure-density gradients are large (Nagao \etal 2011).

 However, excitation of levels lying
very close to each other
implies that the associated cross sections need
to be computed with great accuracy at very low energies in order to
yield reliable rate coefficients. 
The Maxwellian electron distribution at low temperatures
samples only the near-threshold energies above the small excitation
energy of the fine structure transition. Relativistic fine structure
separations therefore
assume special importance even for low-Z atomic ions in determining not
only the energy separation but also the interaction of
the incident electron with the target levels. Owing to its prominence in
astrophysical spectra, a large number of previous studies have been
carried out on electron impact excitation of {\sc O\,iii} (viz. 
compilation of evaluated data by Pradhan and Zhang 2001). Among the recent
ones, whose collision strengths are employed in astrophysical models, are Burke
\etal (1989) and Aggarwal and Keenan (1999).  
But these calculations are basically in LS coupling (Burke \etal
1989), or
with intermediate coupling effects introduced perturbatively via an
algebraic transformation from the LS to LSJ scheme (Aggarwal and Keenan
1999). Although the earlier calculations employed the coupled channel
R-matrix method
which takes account of the extensive resonance structures, the fine
structure separations are not considered. In this report we take account
of both the resonances and fine structure in an ab initio manner.

Another recent development in relativistic
R-matrix codes is the inclusion of the
two-body fine structure Breit interaction terms in the 
Breit-Pauli hamiltonian (Eissner and
Chen 2012, Nahar \etal 2011). A relativistic calculation of collision
strengths can therefore be carried out, including fine structure
explicitly and more accurately than in previous works.
 Relativistic effects are likely to be insignificant for optical
transitions compared to the FIR transitions since the former involve
relatively larger energy separations and relatvistic corrections are small.
Nevertheless, we consider all 10 forbidden transitions among the levels
dominated by the ground configuration of {\sc O\,iii}.
 
     \section{Theory and computations}
A brief theoretical description of the calculations is given.
In particular, we describe relatvistic effects and the representation of
the \eion system.

\subsection{Relativistic fine structure} 
The relativistic Hamiltonian (Rydberg units) in the Breit-Pauli R-matrix (BPRM)
approximation is given by

\begin{equation} 
\begin{array}{l}
H_{N+1}^{\rm BP} = \\ \sum_{i=1}\sp{N+1}\left\{-\nabla_i\sp 2 -
\frac{2Z}{r_i}
+ \sum_{j>i}\sp{N+1} \frac{2}{r_{ij}}\right\}+H_{N+1}^{\rm mass} + 
H_{N+1}^{\rm Dar} + H_{N+1}^{\rm so}.
\end{array}
\end{equation}
where the last three terms are relativistic corrections, respectively:
\begin{equation} 
\begin{array}{l}
{\rm the~mass~correction~term},~H^{\rm mass} = 
-{\alpha^2\over 4}\sum_i{p_i^4},\\
{\rm the~Darwin~term},~H^{\rm Dar} = {Z\alpha^2 \over
4}\sum_i{\nabla^2({1
\over r_i})}, \\
{\rm the~spin-orbit~interaction~term},~H^{\rm so}= Z\alpha^2 
\sum_i{1\over r_i^3} {\bf l_i.s_i}.
\end{array} 
\end{equation}

 Eq.~(2) representes the one-body terms of the Breit interaction. In
addition, another version of BPRM codes including the two-body terms
of the Breit-interaction (Nahar \etal 2011; W. Eissner and G. X. Chen,
in preparation)
has been developed, and is employed in the present work.

\subsection{Effective collision strengths}

 Cross sections or collision strengths at very low energies may be inordinately influenced by
near-threshold resonances. Those, in turn, affect the effective collision
strengths or rate coefficients 
computed by convolving the collision strengths over a Maxwellian
function at a given temperature T as

\be 
\Upsilon_{ij}(T_e) = \int_0^{\infty} \Omega_{ij} (\epsilon)
\exp(-\epsilon/kT_e) d(\epsilon/kT_e), \ee

 where $E_{ij}$ is the energy difference and $\Omega_{ij}$ is the
collision strength for the transition $i \rightarrow j$. The
exponentially decaying Maxwellian factor implies that at low
temperatures only the very low energy $\Omega_{ij}$(E) would
determine the $\Upsilon$(T).

\begin{figure} 
\centering  
\includegraphics[width=\columnwidth,keepaspectratio]{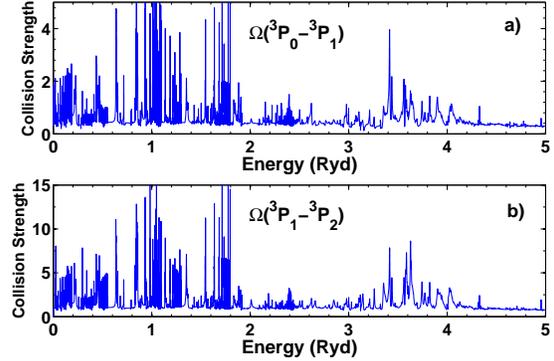}
\caption{Collision strengths for the \o3 IR fine structure transitions
$2p^2 (^3P_0-^3P_1, ^3P_1-^3P_2$) at \lmlm a) 88
\mum and b) 52 \mum respectively. 
High resolution at near-threshold
energies is necessary for accuracy in rate coefficients at low
temperatures. The top panel shows an expanded view in the region E
$\leq$ 1 Rydberg; both transitions have similar resonance
structures. \label{collir}} 
\end{figure}

\begin{figure} 
\centering
\includegraphics[width=\columnwidth,keepaspectratio]{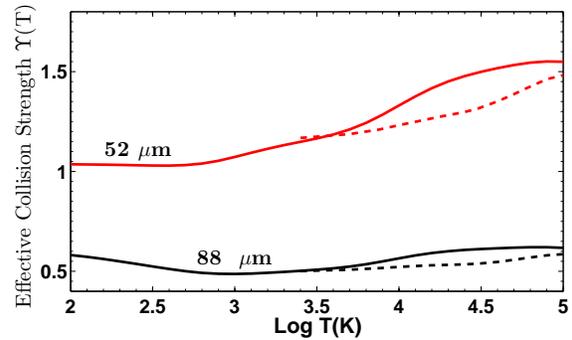}
\caption{Maxwellian averaged effective collision strengths $\Upsilon(T)$
(Eq.~1) for the transitions $^P_0-^3P_1$ at 88 \mum and $^3P_1-^3P_2$ at
52 \mum (solid lines, c.f. Fig.~\ref{collir}).
Previous results without relativistic effects (Aggarwal and Keenan 1999)
are also shown (dashed lines) in the temperature range available \te
$\geq$ 2500 K. \label{upsir}}
\end{figure}

\subsection{Wavefunction representation and calculations}
 Based on the coupled channel approximaton, the R-matrix method (Burke
\etal 1971) entails a wavefunction expansion of the \eion
system in terms of the eigenfuctions for the target ion. In the present
case we are interested in low-lying O/FIR transitions of the ground
configuration $2s^22p^2$. Therefore we
confine ourselves to an accurate wavefunction
representation for the
first 19 levels dominated by the {\it spectroscopic}
 configurations $[1s^2] 2s^22p^2, 2s2p^3, 2s^22p3s$. A much larger set of {\it
correlation} configurations is included for configuration interaction
with the spectroscopic terms using the atomic structure code
SUPERSTRUCTURE (Eissner \etal 1974, Nahar \etal 2003): 
$[1s^2] 2p^4, 2s^22p3p, 2s^22p3d,
2s^22p4s, 2s^22p4p, 2s2p^23s,\\ 2s2p^23p, 2s2p^23d, 2s^23s^2,
2s^23p^2,
2s^23d^2, 2s^24s^2, 2s^24p^2,\\ 2s^23s3p, 2s^23s4s, 2s^23p3d, 2p^33s,
2p^33p, 2p^33d$. 
We note here that the crucial fine structure separations
between the ground state $^3P_{0,1.2}$ levels reproduced
theoretically agree with experimentally measured values to $\sim$3\% (Nahar
\etal 2011, see Pradhan and Nahar (2011) for a general description of
atomic processes and calculations). 
The observed energies were substituted
for theoretical ones in order to reproduce the threshold
energies more accurately. This is of particular importance for
excitation at low temperatures dominated by near-threshold resonances.
Even though the observed and experimental
values are close, a small difference in resonance positions relative to
threshold can introuce a significant uncertainty in the effective
collision strengths. 

 The collision strengths were computed employing the extended Breit-Pauli 
R-matrix (BPRM) codes (Eissner and Chen 2012). Particular care is taken to
test and ensure convergence of collision strengths with respect to
partial waves and energy resolution. Total \eion symmetries up to
(LS)J$\pi$ with J $\leq 19.5$ were included in the calculations, though
it was found that the collision strengths for all forbidden transition
transitions converged for J $\leq 9.5$. An energy mesh of $\Delta E <
10^{-4}$ Rydbergs was used to resolve the near-thresold resonances. The
resonances were delineated in detail prior to averaging over the
Maxwellian distribution. 

\section{Results and Discussion}

 We describe the two main sets of results for the FIR and the 
optical transitions, as well as diagnostics line ratios.

\subsection{Far-infrared transitions}
    The BPRM collision strengths for the two FIR fine structure transitions 
\lmlm 88, 52 \mum are shown in Fig. 1a,b. Although the resonance
structures look similar the magnitude and energy variations are not the
same. The Maxwellian averaged effective collision strengths
$\Upsilon$(T) are quite different, as shown in Fig. 2. While
$\Upsilon(T;^3P_o-^3P_1)$ for the 88 \mum transition 
is relatively constant over three orders of magnitude in temperature, the  
$\Upsilon(T;^3P_1-^3P_2)$ for the 52 \mum transition varies by about a factor
of 1.5 from the low-temperature limit of 100 K to temperatures 
T $>$ 10,000 K. A comparison with the earlier work by Aggarwal and
Keenan (1999) is shown as dashed lines, which range down to their
lowest tabulated temperature 2500K. It can be noted that if the Aggarwal
and Keenan values are extrapolated linearly to lower temperatures then
one would obtain fairly constant effective collision strengths. But the
present results show marked difference owing to resonance structures as
in Fig.~1. 

Such temperature sensitivity of the otherwise density sensitive
52/88 line ratio is illustrated in Fig.~3a,b. In Fig. 3a the solid lines are
ratios with
the present collision strengths, and the dashed lines are using previous
results (Aggarwal and Keenan 1999). We find very good agreement,
implying that at all temperatures down to 2500K the differences in line
ratios would be negligble. 
Fig. 3b shows the
52/88 ratio at various temperatures between 100K and 10,000K. 
Whereas the ratio is
relatively constant with density at 100K, its dependence on density
varies significantly with increasing temperature. The
density-temperature
dianostic value of the 52/88 ratio is apparent from these curves.
Therefore, care must be exercised to establish a temperature
regime for the emitting region. Fig. 3a shows line ratios computed at
2500 K and 10,000K, and variations with electron density. It is also 
found that the values of line ratios at
2500K and 1000K are very close together, implyling covergence for T
$\sim$ 1000K. Fig. 3b clearly demonstrates that
the line ratio decreases rapidly for T $<$ 1000K to almost flat at 100K.
The low-temperature regime 100-1000K is therefore indicated by the
curves shown in Fig.~3b, as well as the limit where the 52/88 ratio is
temperature invariant. So the 52/88 ratio is excellent
density diagnostics in the typical density range Log \ne $\sim$ 3-4 for
T $>$ 1000K without much dependence on temperature (Fig. 3a). 
However, at lower
temepratures the ratio may differ by up to a factor of ten (Fig. 3b). 
Whereas the primary variations are owing to the exponential factors
in $\Upsilon(T)$, (Eq.~3), we emphasize
the role of relativistic fine structure splitting between the
$^3P_{0,1,2}$ levels and near-threshold resonances lying in
between.

\begin{figure} 
\centering
\includegraphics[width=\columnwidth,keepaspectratio]{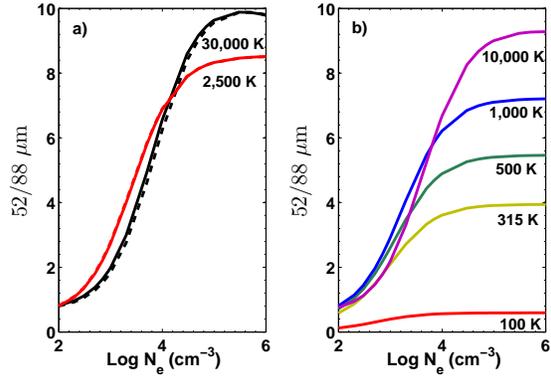}
\caption{The density and temperature dependence of the \lmlm 52/88 \mum line
ratios; a) the solid curves are present results at 2500K and 30000 K and
the dashed lines are using the Aggarwal Keenan (1999) effective collision
strengths; b) 52/88 ratio at temperatures $100 \leq T(K) \leq 10000$.
\label{52/88}}
\end{figure}

\subsection{Forbidden optical transitions}
 Fig. 4 shows the collision strengths for the optical transitions
$^3P_1-^1D_2$, $^3P_2-^1D_2,^1S_0-^1D_2$ at \lmlm 4959, 5007,
4363 respectively. 
 The effective collision strengths for the \o3 optical lines are shown
in Fig.~5. These also
differ significantly from previously available ones, by up to 15\%.
The new results are also obtained down to 100K; their limiting
values at low temperatures tend to 0.4:0.6:1.0. Since \lmlm 4959, 5007
are often blended, it is common to plot the bleneded line ratio
(4959+5007)/4363 shown in Fig.~6. This ratio varies over orders of
magnitude since the upper-most $^1S_0$ level is far less excited at low
energies than the $^1D_2$, and therefore the level populations  and line
intensities depend drastically on temperature. A comparison is made with
fine structure collision strengths derived from the LS term values
of Aggarwal and Keenan (1999) divided according to statistical weights,
again shown as dashed lines in Figs.~5
and 6. However, similar to Fig. 3a), the differences in effective 
collision strengths does
not translate into any significant differences in line ratios even at
Log T = 4.5 ($\approx$ 30000K) wwhere the values differ most.

\begin{figure} 
\centering
\includegraphics[width=\columnwidth,keepaspectratio]{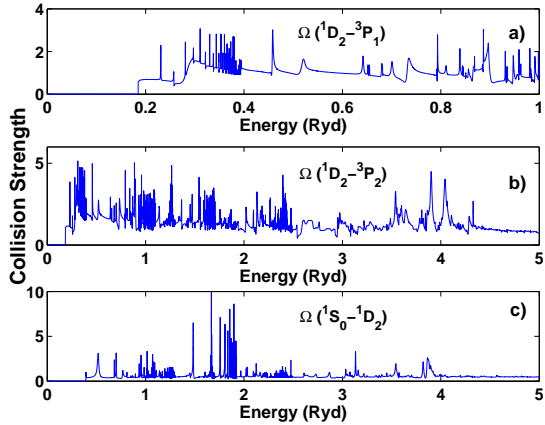}
\caption{Collision strengths of the \o3 optical transitions.
The two transitions $^1D_2-^3P_1, ^1D_2-^3P_2$ in a) and b) 
have similar resonance
structures; the top panel a) presents an expanded view below E $
\leq$ 1 Rydberg.    \label{upsopt}} 
\end{figure}

\begin{figure} 
\centering
\includegraphics[width=\columnwidth,keepaspectratio]{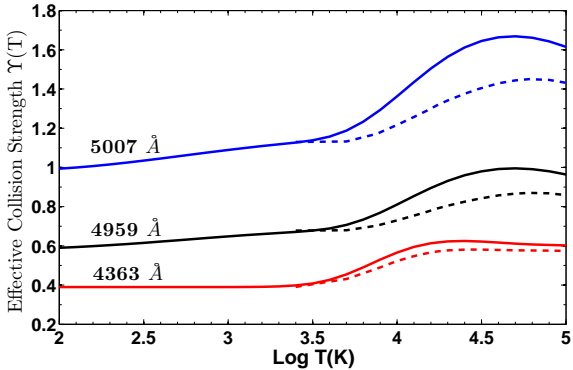}
\caption{Effective collision strengths of the \o3 optical transitions
$^1D_2-^3P_1, ^1D_2-^3P_2, ^1S_0-^1D_2$ at
\lmlm 4959, 5007 and 4363 respectively (c.f. Fig.~4). \label{upsopt}}
\end{figure}

\begin{figure} 
\centering
\includegraphics[width=\columnwidth,keepaspectratio]{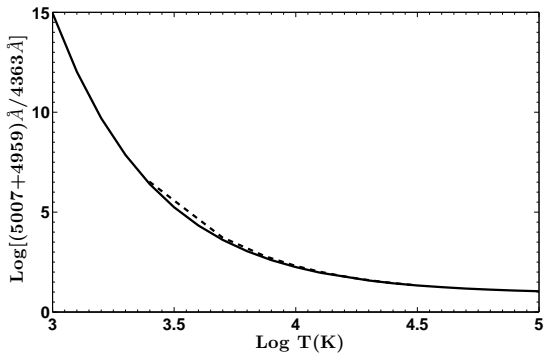}
\caption{Blended \o3 line ratio (4959 + 5007)/4363 vs. temperature,
at \ne = 10$^3$ cm$^{-3}$. The dashed line using earlier data (Aggarwal
and Keenan 1999) is plotted down to 2500K. \label{o3optlr}}
\end{figure}

\subsection{Maxwellian averaged collision strengths}

 In Table 1 we present the effective collision strengths (Eq. 3) for the 10
transitions among the ground configuration levels and their wavelengths.  
The tabulation is carried out at a range of temperatures typical of
nebular environments, including the low temperature range $T \leq 1000$K 
not heretofore considered. 

\begin{table*}
\begin{minipage}{148mm}
\caption{Effective Maxwellian averaged collision strengths}
\begin{tabular}{ccccccccc}
\hline
Transition & $\lambda$ & $\Upsilon$(100) & $\Upsilon$(500) &
$\Upsilon$(1000) & $\Upsilon$(5000) & $\Upsilon$(10000) & 
$\Upsilon$(20000) & $\Upsilon$(30000)\\
\hline
 $^3P_0-^3P_1$ & 88 $\mu$m & 5.814(-1) & 5.005(-1) & 4.866(-1) & 5.240(-1)
& 5.648(-1) & 6.007(-1) & 6.116(-1)\\
 $^3P_2-^3P_0$ & 33 $\mu$m & 2.142(-1) & 2.153(-1) & 2.234(-1) &
2.469(-1) & 2.766(-1) &	3.106(-1) & 3.264(-1)\\
 $^3P_2-^3P_1$ & 52 $\mu$m & 1.036(0) & 1.032(0) & 1.072(0) &
1.210(0) & 1.330(0) & 1.451(0) & 1.499(0)\\
 $^1D_2-^3P_0$ & 4933 $\AA$ & 1.959(-1) &	2.088(-1) & 2.154(-1) &
2.347(-1) & 2.693(-1) & 3.094(-1) & 3.256(-1)\\ 
 $^1D_2-^3P_1$ & 4959 $\AA$ & 5.903(-1) &	6.285(-1) & 6.483(-1) &
7.067(-1) & 8.108(-1) &	9.313(-1) & 9.802(-1)\\
 $^1D_2-^3P_2$ & 5007 $\AA$ & 9.934(-1) &	1.056(0) & 1.089(0) &
1.188(0) & 1.363 (0) &	1.564(0) & 1.645(0)\\
 $^1S_0-^1D_2$ & 4363 $\AA$ & 3.900(-1) &	3.899(-1) & 3.899(-1) &
4.544(-1) & 5.661(-1) &	6.230(-1) & 6.219(-1)\\
 $^1S_0-^3P_1$ & 2321 $\AA$ & 1.765(-1) &	1.590(-1) & 1.477(-1) &
1.228(-1) & 1.223(-1) &	1.294(-1) & 1.332(-1)\\
 $^1S_0-^3P_2$ & 2332 $\AA$ & 2.850(-1) &	2.587(-1) & 2.421(-1) &
2.045(-1) & 2.046(-1) &	2.170(-1) & 2.235(-1)\\
 $^1S_0-^3P_0$ & 2317 $\AA$ & 5.965(-2) &	5.354(-2) & 4.959(-2) &
4.094(-2) & 4.069(-2) &	4.299(-2) & 4.424(-2)\\
\hline
\end{tabular}
\end{minipage}
\end{table*}

\subsection{Conclusion}

 Improved collision strengths including fine structure with relativistic
effects are computed. Owing to the diagnostic 
importance of the \o3 forbidden FIR and optical lines, the relatively
small but significant differences of up to 20\% should provide more
accurate line ratios. Particular attention is paid to the resolution of
resonances in the very small energy region above threshold(s), enabling
the study of low temperature behavior.

 The line emissivities and ratios computed in this work
demonstrate the temperature-density behaviour at low
temperatures and at typical nebular temperatures. However, depending on
the astrophysical sources a complete
model of line emissivities may also need to take into consideration the
Bowen fluorescence mechanism: the {\it radiative}
 excitation of $2p^2 \ ^3P_2 - 2p3d
^3P^o_2$ by He~{\sc ii} Ly$\alpha$ at 304$\AA$ and cascades into
the upper levels of the forbidden transitions considered herein
(viz. Saraph and Seaton 1980, Pradhan and Nahar 2011). 
In addition, for higher
temperatures T $>$ 20,000K proton impact excitation of the ground state
fine structure levels $^3P_{0,1,2}$ needs to be taken into account;
at lower temperatures the excitation rate coefficent due
to electrons far exceeds that due to protons (Ryans \etal 1999).
Finally, there may be some contribution from \eion recombination from
{\sc O\,iv} to \o3, since recombination rate coefficients increase
sharply towards lower temperatures while collisional excitation rates
decrease (level-specific and total recombination rate coefficients 
may be obtained from S. N. Nahar's database NORAD at:
www.astronomy.ohio-state.edu\$sim$nahar$). Recombination contribution depends on
the {\sc O\,iv}/\o3 ionization fraction, which at low temperatures
would be small.

\section*{Acknowledgments}
 The computational work was 
carried out at the Ohio Supercomputer Center in Columbus 
Ohio. This work was partially supported by a grant from the NASA
Astrophysical Research and Analysis program. EP would like to
gratefully acknowledge a Summer Undergradute Research Program grant from the
Ohio State University.

\label{lastpage}


\begin{thebibliography}{}
\frenchspacing
\def\aa{{\it Astron. Astrophys.}\ }
\def\aasup{{\it Astron. Astrophys. Suppl. Ser.}\ }
\def\adndt{{\it Atom. data and Nucl. Data Tables.}\ }
\def\aj{{\it Astron. J.}\ }
\def\apj{{\it Astrophys. J.}\ }
\def\apjs{{\it Astrophys. J. Supp. Ser.}\ }
\def\apjl{{\it Astrophys. J. Lett.}\ }
\def\baas{{\it Bull. Amer. Astron. Soc.}\ }
\def\cpc{{\it Comput. Phys. Commun.}\ }
\def\jpb{{\it J. Phys. B}\ } 
\def\jqsrt{{\it J. Quant. Spectrosc. Radiat. Transfer}\ }
\def\mn{{\it Mon. Not. R. astr. Soc.}\ }
\def\pasp{{\it Pub. Astron. Soc. Pacific}\ }
\def\pra{{\it Phys. Rev. A}\ }
\def\pr{{\it Phys.  Rev.}\ } 
\def\prl{{\it Phys. Rev. Lett.}\ }

\bibitem{} Aggarwal, K. M. and Keenan, F. P., 1999, \apjs, 123, 311 
\bibitem{} Aller, L. H. 1956, {\it Gaseous Nebulae}, Wiley, New York
\bibitem{ben95} Berrington K.\,A., Eissner W. \& Norrington P.\,H., 1995,
Comput. Phys. Commun. 92, 290
\bibitem{} Burke, V. M., Lennon, D. J., and Seaton, M. J., 1989, \mn,
236, 353 
\bibitem{} Crawford, F. J., Keenan, F. P., Aggarwal, K. M., Wickstead,
A. W., Aller, L. H. and Feibelman, W. A., 2000, 362, 730
\bibitem{dp03} Dopita M.\,A. \& Sutherland R.\,S., 2003, {\it Astrophysics
of the Diffuse Universe}, Springer-Verlag
\bibitem{es72} Eissner W. \& Seaton M.\,J., 1972, J. Phys. B 5, 2187
\bibitem{} Houck, J. R. \etal, 2004, \apjs, 154, 18
\bibitem{} Houck, J. R. \etal, 2005, \apjl, 622, L105
\bibitem{} Liu, X., Shapley, A. E., Coil, A. L., Brenchmann, J. and Ma
C.-P., 2008, \apj, 678, 758
\bibitem{} Martin-Hernandez \etal, 2002, \aa, 381, 606
\bibitem{} Morisset, C., Schaerer, D., Martin-Hernandez, N. L., Peeters,
E., Damour, F., Baluteau, J.-P., Cox, P., and Roelfsema, P., 2002, \aa,
386, 558
\bibitem{} Nahar, S. N., Eissner, W., Chen, G.-X., Pradhan, A. K., 2003,
408, 789
\bibitem{} Nahar, S. N., Pradhan, A. K., Montenegro, M., Eissner, W.,
2011, \pra, 83, 053417
\bibitem{} Nagao, T., Maiolino, R., Marconi, A., and Matsuhara, H. 2011,
\aa, 526, A149
\bibitem{} Pradhan, A. K. and Nahar, S. N., 2011, {\it Atomic
Astrophysics and Spectroscopy}, Cambridge University Press
\bibitem{} Pradhan, A. K. and Zhang, H. L., Landolt-B\"{o}rnstein
Volume 17 {\it Photon and Electron Interactions
with Atoms, Molecules, Ions}, Springer-Verlag, 2001, 
(Ed: Y. Itikawa), I.17.B, 1
\bibitem{} Ryans, R. S. I., Foster-Woods, V. J., Reid, R. H. G., Keenan,
F. P., 1999, \aa, 345, 663
\bibitem{} Saraph, H. E. \& Seaton, M. J., 1980, \mn, 193, 617

\end{thebibliography}
\end{document}